\documentclass[twocolumn]{rps-esrel2019}

\usepackage{amsmath,amssymb,amsfonts}
\usepackage{algorithmic}
\usepackage{graphicx}
\usepackage{textcomp}
\usepackage{xcolor}
\usepackage[rflt]{floatflt}
\usepackage[caption=false]{subfig}

\usepackage{tikz}
\usetikzlibrary{automata,backgrounds,calc,fit,positioning,shapes}
\usepackage{pgfplots}
\usepgfplotslibrary{fillbetween}
\usepackage[overlay,absolute]{textpos}
\usepackage{transparent}

\def\papername{\jobname}

\usepackage[square,sort,comma,numbers]{natbib}

\usepackage[color=white,disable]{todonotes}

\usepackage{multirow}

\g@addto@macro\UrlBreaks{\do\-}
\newcommand{\simulink}{\textsc{Simulink}}

\newcommand{\prism}{\textsc{Prism}}
\newcommand{\errorpro}{\textsc{OpenErrorPro}}
\newcommand{\more}{\textsc{MoRe}}
\newcommand{\simpars}{\textsc{SimPars}}

\begin{document}

\markboth{Clemens Dubslaff, Kai Ding, Andrey Morozov, Christel Baier, and Klaus Janschek}{
Breaking the Limits of Redundancy Systems Analysis}

\twocolumn[

\title{Breaking the Limits of Redundancy Systems Analysis}

\author{Clemens Dubslaff, Kai Ding, Andrey Morozov, Christel Baier, and Klaus Janschek}

\address{Technische Universit\"at Dresden, Germany.\\
\email{\{clemens.dubslaff, kai.ding, andrey.morozov, christel.baier, klaus.janschek\}@tu-dresden.de}}

\begin{abstract}
Redundancy mechanisms such as triple modular redundancy protect
safety-critical components by replication and thus improve systems fault tolerance.
However, the gained fault tolerance comes along with costs to be invested,
e.g., increasing execution time, energy consumption, or packaging size,
for which constraints have to be obeyed during system design.
This turns the question of finding suitable combinations of components to be protected
into a challenging task as the number of possible protection combinations grows exponentially
in the number of components.
We propose \emph{family-based approaches} to tackle the combinatorial blowup 
in redundancy systems modeling and analysis phases. Based on systems designed in 
$\simulink$ we show how to obtain models that include all possible protection combinations
and present a tool chain that, given a probabilistic error model, 
generates discrete Markov chain families. 
Using symbolic techniques that enable concise family representation
and analysis, we show how $\simulink$ models of realistic size can be protected
and analyzed with a single family-based analysis run while a one-by-one 
analysis of each protection combination would clearly exceed any realistic
time constraints.
\end{abstract} 

]
\vspace*{-2em}
\section{Introduction}

Fault tolerance plays a significant role in the design of safety-critical systems 
as it enables a system to continue functioning in the presence of faults.
The key underlying technique to achieve fault tolerance is provided by \emph{redundancy}, 
i.e., mechanisms to discover and evaluate faults depending on the behaviors 
of replicated system components.
For instance, components can be protected by \emph{triple modular redundancy (TMR)} 
where the component is triplicated and their results are processed through
a majority voting mechanism into a single output.
Though protecting components reduces the overall probability of failure, it increases
costs in terms of, e.g., the systems packaging size, energy consumption, 
production costs, or execution time. These costs hinder to apply the naive approach 
of protecting all system components to maximize reliability and motivate the task of
protecting only some components towards a good tradeoff between reliability and costs. 
However, the number of possible protection combinations grows exponentially in the
number of components, which renders the design of redundancy systems a challenging task:
Systems designers might have to model and analyze a huge amount of protection 
combinations before they find a combination that achieves a satisfactory tradeoff. 
As even the (tradeoff) analysis of a single protection combination can take
a significant amount of time, this iterative development cycle is likely to exceed 
time constraints. Furthermore, one might be not interested in a protection
combination that is only satisfactory but optimal with respect to the tradeoff,
rendering an exhaustive analysis of all combinations inevitable.

To tackle the aforementioned challenges when modeling and analyzing redundancy systems, 
we propose to use \emph{family-based approaches}~(see, e.g., \cite{Thum14,DBK15,CDKB16,BD18})
where a single \emph{family model} comprises the behaviors of all protection combinations. 
First, such approaches avoid modeling each protection combination individually
and to use an automated generation of any combination out from the family model.
Thus, the design process of redundancy systems turns from alternating modeling
and analyzing phases to a single modeling phase followed by an analysis phase
those results yield suitable protection combinations.
Second, family-based approaches enable an \emph{all-in-one analysis} where the family
model is analyzed in a single run instead of analyzing each family member
in isolation. This allows for exploiting commonalities between the family members
using symbolic representations and analysis operations, e.g., by 
\emph{binary decision diagrams} (BDDs, cf.~\cite{McMillan,BK08}).
In the feature-oriented systems domain~\cite{Thum14}, such symbolic
techniques have shown drastic speedups for quantitative analyses~\cite{DBK15} 
especially when family members share lots of behaviors. 
As redundancy mechanisms introduce several identical components in the system, 
redundancy systems are naturally eligible for such a concise model 
representation and analyses using symbolic methods. 
Especially when one is interested in \emph{optimal} protection combinations for 
which an exhaustive analysis of all family members can hardly be avoided, 
such an all-in-one analysis might mitigate the limits induced by the
combinatorial blow-up in the number of system components.

We demonstrate the benefits of family-based approaches for the modeling and 
analysis of redundancy systems by a tool chain where redundancy mechanisms
are introduced in $\simulink$ models and a family-based analysis is 
performed on discrete Markov chain (DTMC) families using the symbolic
probabilistic model checker $\prism$~\cite{prism40}.
Our $\simulink$ models with redundancy
are obtained by an annotative approach where $\simulink$ blocks
following the model-based redundancy technique ($\more$, cf. \cite{more}).
Specifically, we consider $\simulink$ redundancy design patterns 
such as \textit{comparison}, \textit{voting} (i.e., TMR), and \textit{sparing}
that replace $\simulink$ blocks amendable for protection in combination with a probabilistic 
error model. For the automated generation of the DTMC family model, we employ $\simpars$ and 
$\errorpro$~\cite{morozov2016stochastic}.

As illustrative case studies we issue two control loops modeled in $\simulink$:
a proportional-integral-derivative (PID) controller and a
velocity control loop (VCL) of an aircraft velocity model borrowed from the 
$\simulink$ example set~\cite{aircraft}. While the PID family comprises
64 members with comparably small model sizes %
the VCL has 65\,536 family members those state space exceeds $10^{11}$ states,
making symbolic methods for their analysis inevitable.
First, we synthesize protection combinations in the PID example 
that are optimal with respect to tradeoffs expressed through 
\emph{quantiles}~\cite{BDKDKSW14,BDKL14}, i.e.,
maximizing the number of control loop rounds where the probability of
failure is guaranteed to be below a given threshold. This example shows
that our approach can be used to investigate properties that can hardly
be analyzed using de-facto standard simulative approaches.
Second, we determine protection combinations in the VCL model that
are Pareto-optimal with respect to the probability of failure within two rounds 
of the control loop and its execution time, solving the following problems:
\begin{enumerate}
	\item minimize the probability of failure while staying within a given execution time, and
	\item minimize execution time while not exceeding a certain probability of failure.
\end{enumerate}
We show that for the VCL model an all-in-one
analysis gains a speedup in three orders of magnitude compared to the
one-by-one analysis. In particular, the presented all-in-one analysis manages 
to obtain results in less than 5~hours while a one-by-one analysis would 
require around 250~days of computing time, clearly exceeding acceptable 
time constraints in systems design.

\paragraph{Outline.}
Section~\ref{sec:relatedwork} discusses
related work and techniques used in the paper. The general approach 
towards $\simulink$ family models with redundancy and their translation
into DTMC families is described in Section~\ref{sec:approach_PID}. 
In Section~\ref{sec:breakinglimits} the analysis of PID and VCL families
is carried out and optimal protections are synthesized.
We close our paper and discuss further work in Section~\ref{sec:conclusion}.

\section{Related Work and Concepts}\label{sec:relatedwork}

\paragraph{Probabilistic Model Checking.}
For the analysis of Markovian stochastic models, probabilistic model 
checking (PMC, cf.~\cite{BK08}) is an automated technique 
that has been successfully applied to numerous real-world case
studies to analyze systems performance and Quality of Service.
We rely on \emph{discrete Markov chains (DTMCs)} as stochastic model, 
i.e., state-transition graphs where the transitions are purely probabilistic.
\emph{Symbolic methods} can compete with the well-known state-space explosion
problem by concise model representations, e.g., through \emph{binary decision diagrams} 
(BDDs, cf.~\cite{McMillan,BK08}). 
The prominent PMC tool $\prism$~\cite{prism40} uses 
\emph{multi-terminal BDDs}~\cite{FMcGY97,BFGHMPS97} 
for a purely symbolic analysis without the use of an enumerative model representation. 
It is well-known that the size of symbolic representations by BDDs is sensitive 
to the so-called \emph{variable order}.
In~\cite{KBCDDKMM16}, variable-reordering techniques towards a compact state-space
representation have been introduced for $\prism$. This enabled the analysis of 
large-scale systems and speedup their analysis, e.g.,  
for the all-in-one family-based analysis of feature-oriented systems~\cite{DBK15,CDKB16}.
Family-based synthesis using symbolic PMC towards optimal 
system configurations has been detailed in \cite{BD18}.

\paragraph{Reliability Analysis and Variability in $\simulink$.}
Reliability analysis of $\simulink$ using verification techniques has been considered 
in~\cite{Joshi2005}, where a handcrafted tool chain from $\simulink$ to the 
programming language \textsc{Lustre} in combination with the \textsc{Scade}
design verifier has been used. The \textsc{ModIFI} approach presented 
in~\cite{svenningsson2010modifi} implements fault-injection capabilities in
$\simulink$ to evaluate error handling mechanisms. However, their error model
is non-probabilistic and their analysis method is based on simulations --
an extension of this work towards a stochastic reliability analysis thus
is not as easy. In \cite{Beer2013ModelBasedQS} a transformation of $\simulink$
models to continuous-time Markov chains is presented, used to perform 
dependability analysis using the model checker $\prism$. 
Variability modeling in $\simulink$ has been considered in \cite{Haber2013},
using a direct implementation of a delta-oriented approach.
As we will show, we do not have to rely on this powerful formalism
to model families of redundancy systems. However, their approach
could possibly be combined towards a family-based analysis
of a great variety of $\simulink$ models.

\paragraph{DEPM and $\errorpro$.}
The \emph{Dual-graph Error Propagation Model}~(DEPM, \cite{morozov2015errorpro})
comprises a state-based model of the control- and data-flow,
both linked to a stochastic error model.
In contrast to (static) fault-tree analysis~\cite{Mattarei16}, DEPMs can capture 
recurrent and hierarchical behaviors. 
$\errorpro$ is a tool that provides the automated generation from DEPMs to its
DTMC semantics in terms of $\prism$ code, enabling the analysis of DEPMs for
manifold properties by a great variety of tools.
Also due to $\errorpro$'s support of many base-line formalisms that can be translated
to DEPMs, they are well-suited for the reliability analysis of safety-critical systems.
For instance, $\simulink$ models can be translated to DEPMs using $\simpars$~\cite{morozov2016stochastic}.

\paragraph{Further Methods in Reliability Analysis.}
Simulation-based approaches are the usual method for reliability analysis.
However, as faults are usually \emph{rare events} with a relatively 
small probability, sufficient confidence in analysis results
are difficult to achieve. We are aware of techniques that could
circumvent these issues (see, e.g., \cite{LegayST16} for an overview) 
but their implementation would have went far behind the purpose
of this paper and such approaches would not allow for a tradeoff
analysis in terms of quantiles.

Symbolic techniques for the reliability analysis of safety-critical
systems have been first and foremost applied to tackle the state-space
explosion system, e.g., in the field of fault-tree analysis 
(see, e.g., \cite{Mattarei16}). We are not aware of any application
that uses such techniques also to compete with the combinatorial
blowup in the number of system configurations such as protection 
mechanisms we deal with in this paper.

\section{From $\simulink$ to DTMC Families}\label{sec:approach_PID}
The core of our approach towards a family-based analysis of $\simulink$
models with various protection combinations lies in the generation
of $\simulink$ models with redundancy that can be transformed
into DTMC families. We illustrate the modeling and the step by step 
transformation using the $\simulink$ model of a 
proportional-integral-derivative (PID) controller. 
A PID is one of the most important and widely used
feedback controllers to apply accurate and responsive correction 
to a control function, essential for many industry areas such as aerospace, 
process control, manufacturing, and robotics.
\begin{figure}[ht]
	\includegraphics[width=\linewidth]{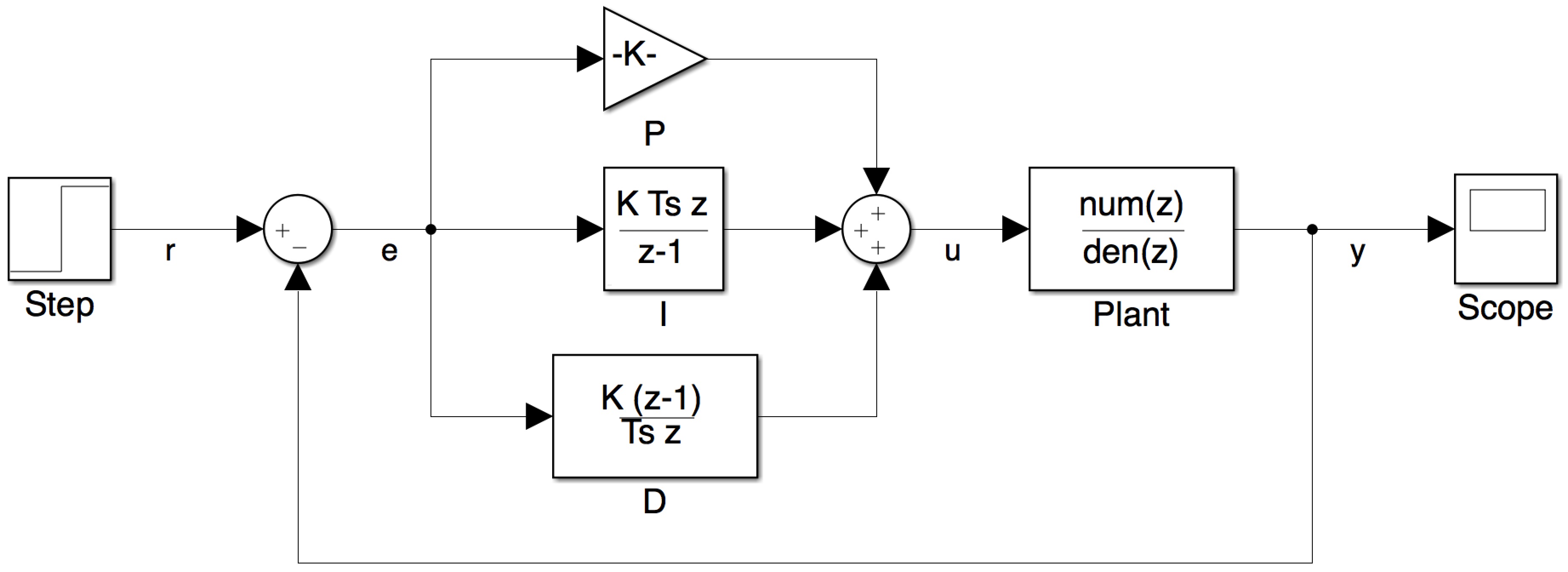}
	\caption{\label{fig:pid} PID controller is designed $\simulink$ with separate blocks for the P, I, and D terms.}
\end{figure}
Figure~\ref{fig:pid} shows the $\simulink$ PID controller 
using separate blocks for the P, I, and D terms.

\begin{figure*}[ht]
	\centering
	\subfloat[comparison]{\includegraphics[width=.31\textwidth]{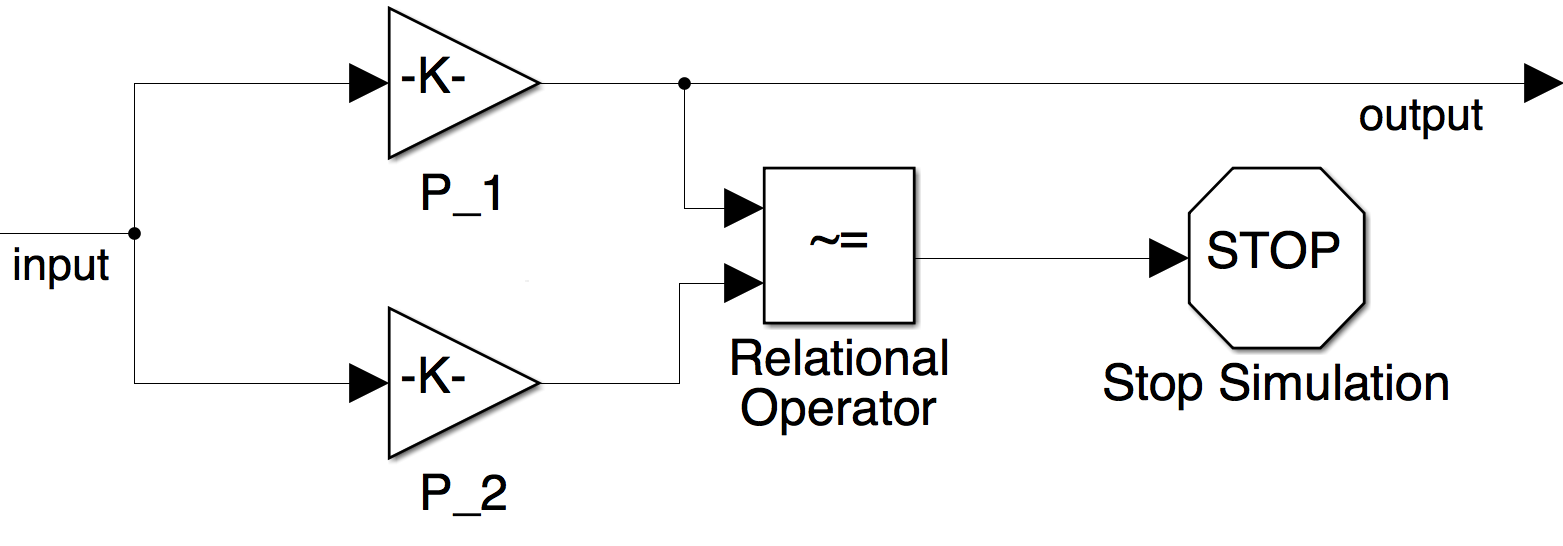}}\hspace{1em}
	\subfloat[voting]{\includegraphics[width=.31\textwidth]{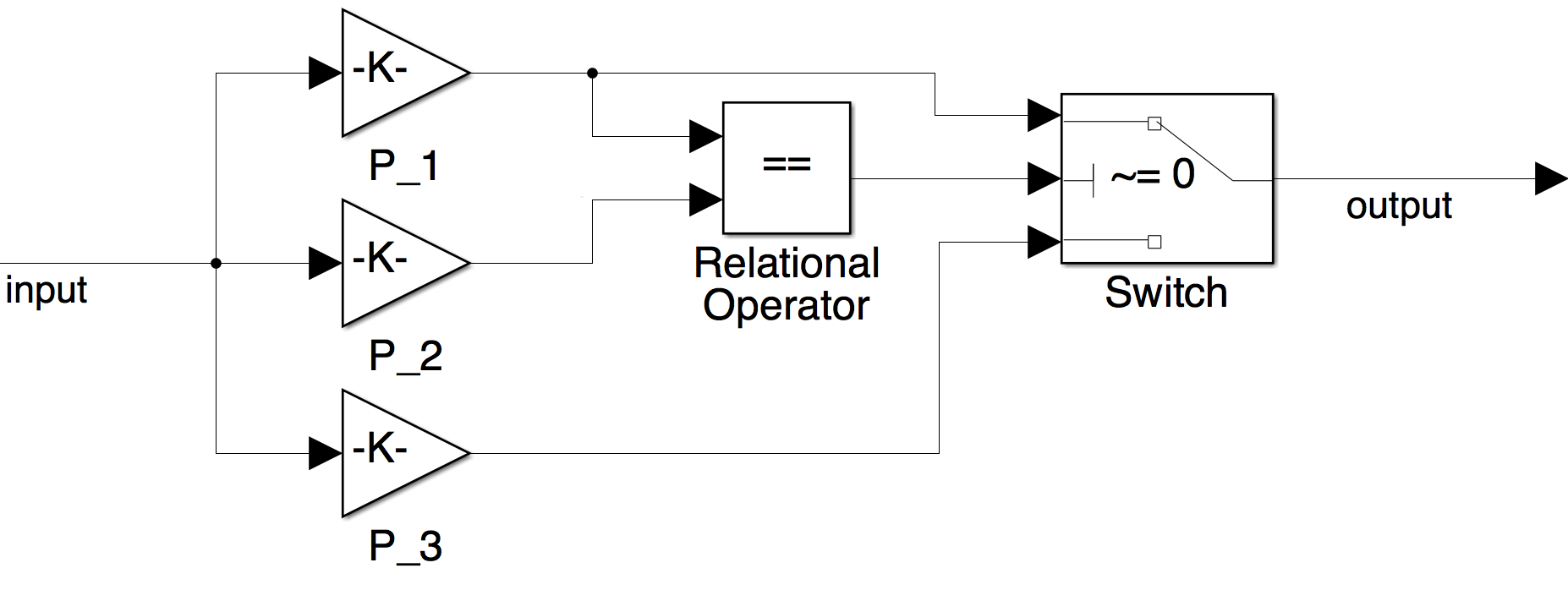}}\hspace{1em}
	\subfloat[sparing]{\includegraphics[width=.31\textwidth]{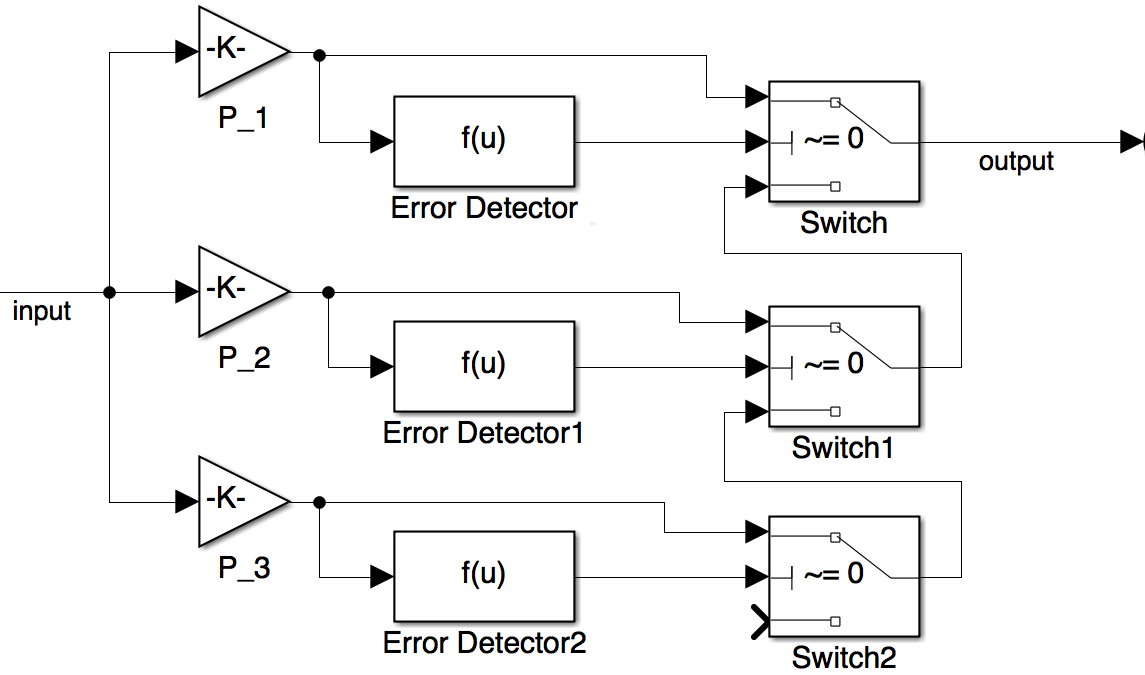}}
	\caption{\label{fig:redundancy} Redundancy mechanisms for $\simulink$ blocks, 
		illustrated for a P element}
\end{figure*}

\subsection{Protecting $\simulink$ Blocks}\label{ssec:protections}
To introduce redundancy mechanisms into $\simulink$ models, we consider
syntactic transformation rules that describe how to obtain protected 
blocks from non-protected ones.
Specifically, we consider here the following three redundancy mechanisms:
\begin{description}
	\item[(comparison)] The block is duplicated and both outputs are compared.
		In case their output differs a dedicated failure state is reached. 
		Otherwise, the output is the one of both blocks.
	\item[(voting)] Following the triple-modular-redundancy principle (TMR),
		the block is triplicated and the output is based on
		a majority decision.
	\item[(sparing)] One block is operational and the remaining two blocks serve as spares.
		If an error in an active block is detected by a built-in error detection unit, 
		a spare block takes over.	
\end{description}
Figure~\ref{fig:redundancy} illustrates how to obtain protections from a given
block, exemplified by the P term of the PID controller.
Note that these patterns are modeled in such a way that that they share most of the
behaviors by using multiple times the original block, %
compare block, and switch block.

\subsection{$\simulink$ Models with Redundancy}\label{ssec:simulinkredundancy}
\begin{figure*}[hbtp]
	\includegraphics[width=\textwidth]{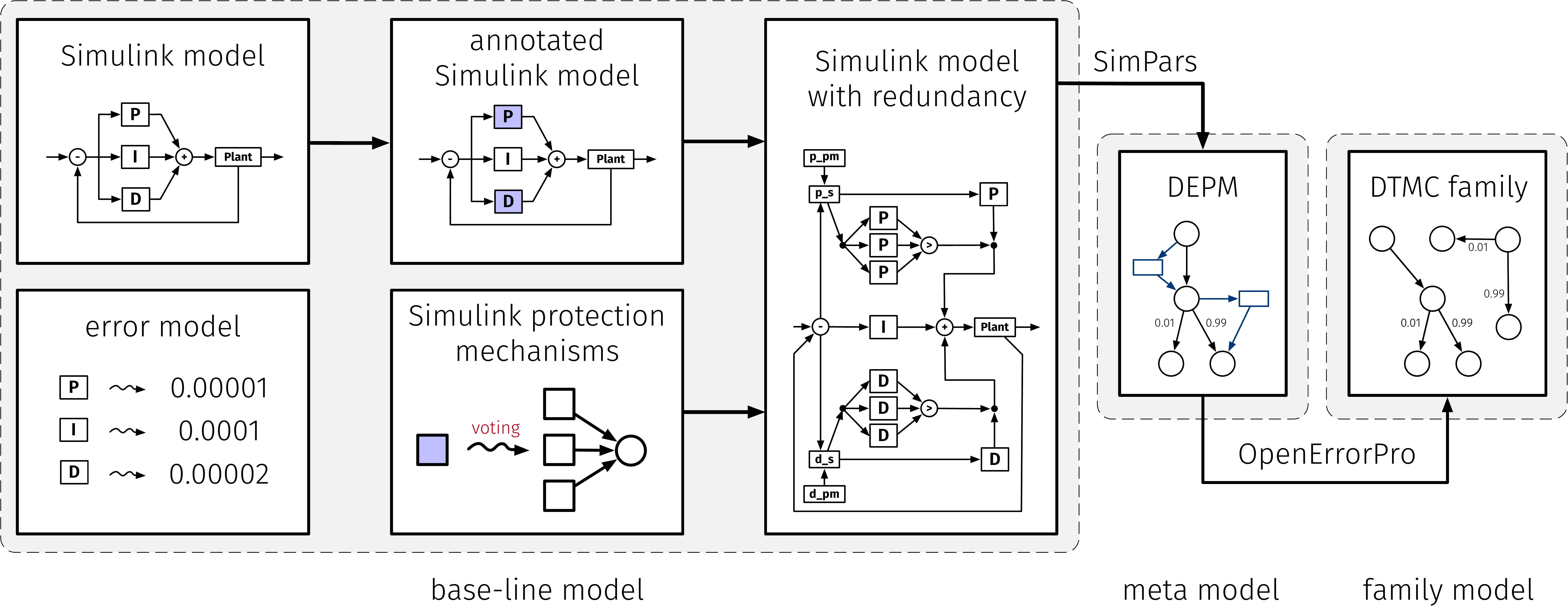}
	\caption{\label{fig:schema} Schema of the approach, obtaining DTMC 
		families from annotated $\simulink$ models}
\end{figure*}

The general workflow of our approach towards a DTMC family and their
analysis out from $\simulink$ models is depicted in Figure~\ref{fig:schema}
using the PID example.
First, blocks in the base-line model are annotated with the types 
of protections that should be considered, e.g., comparison, voting, and sparing.
In Figure~\ref{fig:schema}, we annotated voting protections for the P and D term 
in the PID example (indicated by shaded blocks). 
Then, the annotated base-line model is transformed to
a $\simulink$ model that includes redundancies by replacing every block \textbf{x}
with a switch block \textbf{x\_s} those purpose is to select the 
protection mechanism, followed by the redundancy blocks illustrated in the
Figure~\ref{fig:redundancy} according to the type annotations. 
The switch depends on a free variable \textbf{x\_pm}
that stands for the kind of protection chosen -- the output of the switch connect
to the inputs of the selected mechanism \textbf{x\_pm}.
Thus, by setting the variable \textbf{x\_pm}, e.g., to ``no'' or
``voting'', the control flow of the model either follows no protection
or the voting protection, respectively. The final $\simulink$ model with redundancy 
resulting from the PID controller possibly protecting the P and D term with 
the voting pattern is depicted in the center of Figure~\ref{fig:schema}.
Note that this model stands for a family of controllers comprising four
family members that can be selected by setting variables \textbf{p\_pm} and
\textbf{d\_pm} to either ``no'' or ``voting''.
Hence, a redundancy systems designer does not have to model each protection
combination in isolation but only has to specify the syntactic protection
rules and annotate blocks for protection -- the resulting $\simulink$
model with the desired protections can easily be selected through
choosing the switch variables.

\subsection{DTMC Families}
Having obtained the $\simulink$ model with redundancy that stands for a
family of models with different protection combinations, we use $\simpars$
to generate a Dual-graph Error Propagation Model (DEPM)~\cite{morozov2016stochastic}
preserving the switch variables \textbf{x\_pm} as data elements. 
For this, we employ an error model that assigns to each $\simulink$ block 
the probability for some fault occurring in this block. This
simple error model could also imagined 
to involve further $\simulink$ blocks or statistical data about faults.
In the last step towards DTMC families, $\prism$ code representing 
the family of DTMCs that models the control flow and fault propagation 
is automatically generated using the tool $\errorpro$~\cite{morozov2015errorpro}. 
Also in this step, switch variables are maintained such
that also single family members can be extracted from the DTMC family member
by choosing protection mechanisms in the switch variables of the $\prism$ model. %
\newcommand{\tof}{{\color{red}\textbf{?!?}}}
\newcommand{\res}[4]{#1&#2&#3&#4}
\begin{table*}[h]
	\tbl{\label{tab:pidpfail} PID analysis results for \textbf{(pfail)} with $n=10$ and \textbf{(qround)} with $\theta=3{\cdot}10^{-4}$}
	{\tabcolsep5pt 	
	\begin{tabular}{@{}r|cccc|cccc|cccc|cccc@{}}\toprule 
		&\multicolumn{4}{c|}{D:--}&\multicolumn{4}{c|}{D:c}&\multicolumn{4}{c|}{D:v}&\multicolumn{4}{c}{D:s}\\
		I&P:--&P:c&P:v&P:s&P:--&P:c&P:v&P:s&P:--&P:c&P:v&P:s&P:--&P:c&P:v&P:s\\\colrule
		--&\res{$6.1$}{5.2}{5.2}{6.1}&\res{3.5}{2.6}{2.6}{3.5}&\res{3.5}{2.6}{2.6}{3.5}&\res{$6.1$}{5.2}{5.2}{6.1}\\
		c&\res{4.4}{3.5}{3.5}{4.4}&\res{1.8}{0.9}{0.9}{1.8}&\res{1.8}{0.9}{0.9}{1.8}&\res{4.4}{3.5}{3.5}{4.4}\\
		v&\res{4.4}{3.5}{3.5}{4.4}&\res{1.8}{0.9}{0.9}{1.8}&\res{1.8}{0.9}{0.9}{1.8}&\res{4.4}{3.5}{3.5}{4.4}
		\\\multirow{-4.2}{*}{\begin{sideways}\textbf{(pfail)} in $10^{-5}$\end{sideways}}\hspace{1em}s&\res{$6.1$}{5.2}{5.2}{6.1}&\res{3.5}{2.6}{2.6}{3.5}&\res{3.5}{2.6}{2.6}{3.5}&\res{$6.1$}{5.2}{5.2}{6.1}\\\colrule
		--&\res{43}{50}{50}{43}&\res{75}{100}{100}{73}&\res{75}{100}{100}{75}&\res{43}{50}{50}{43}\\
		c&\res{60}{75}{75}{60}&\res{150}{300}{300}{151}&\res{150}{300}{300}{151}&\res{60}{75}{75}{60}\\
		v&\res{60}{75}{75}{60}&\res{149}{299}{299}{151}&\res{149}{299}{299}{151}&\res{60}{75}{75}{60}
		\\\multirow{-4}{*}{\begin{sideways}\textbf{(qround)}\end{sideways}}\hspace{1em}s&\res{43}{50}{50}{43}&\res{75}{100}{100}{73}&\res{75}{100}{100}{75}&\res{43}{50}{50}{43}\\\botrule
	\end{tabular}}
\end{table*}
\begin{table*}[ht]
	\tbl{\label{tab:analysis}Statistics to the analysis experiments}
	{\tabcolsep4pt
	\begin{tabular}{@{}l|ll|c|cc|cc@{}}\toprule 
		case study&property&parameter&states&\multicolumn{2}{c|}{all-in-one analysis}&
		\multicolumn{2}{c}{one-by-one analysis}\\
		&&&&nodes&time [s]&$\Sigma$ nodes&time [s]\\\colrule
		PID&\textbf{(pfail)}&$n=10$&6\,781\,782&16\,935&12.9&240\,946&37.8\\
		&\textbf{(qround)}&$\theta=0.0003$&"&"&6\,693.1&"&2\,531.4\\\colrule
		VCL&\textbf{(pfail)}&$n=2$&$4.7{\cdot}10^{13}$&1\,949\,466&17\,344.1&$\approx\!1.2{\cdot}10^{12}$&$\approx\!2.2{\cdot}10^7$\\
	\end{tabular}}
\end{table*}
\begin{figure*}[hbtp]
	\includegraphics[width=.95\textwidth]{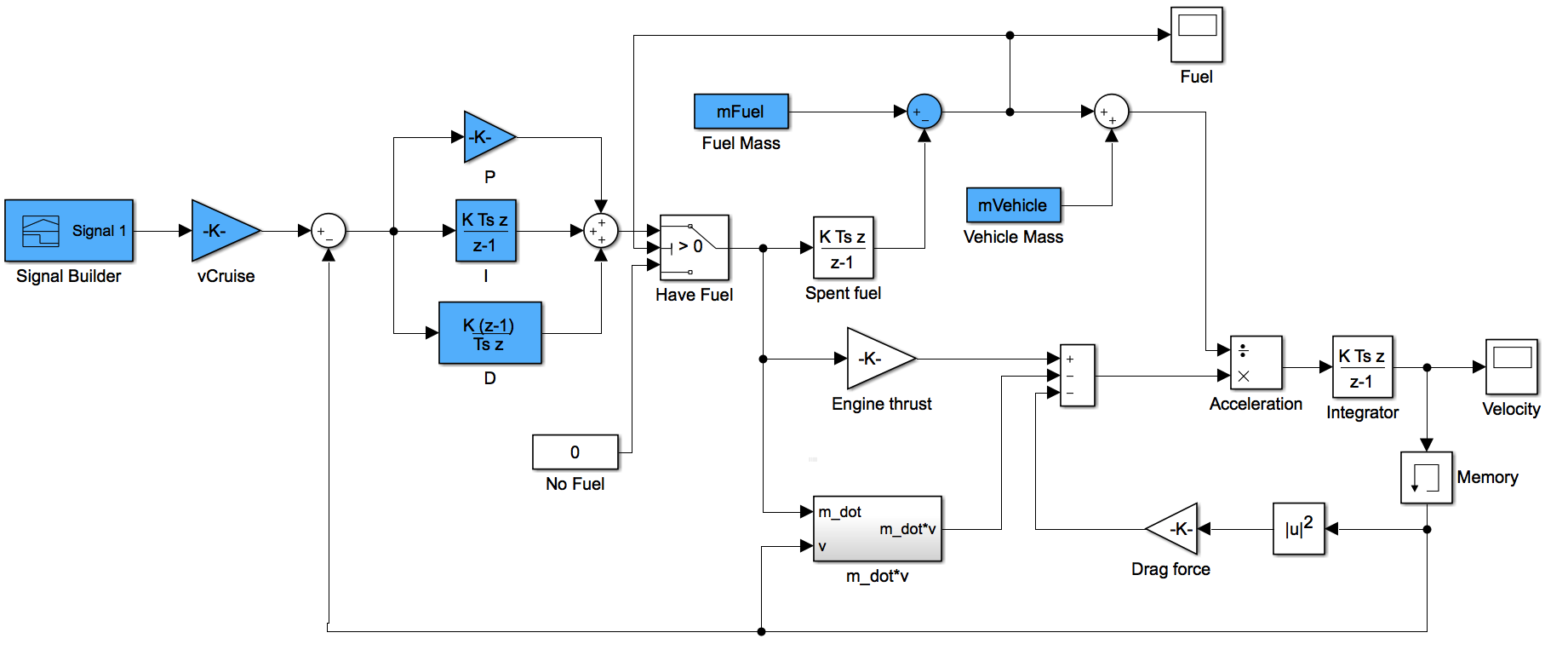}
	\caption{\label{fig:velocity} The $\simulink$ aircraft velocity control loop (VCL) model,
		blocks to be protected highlighted}
\end{figure*}
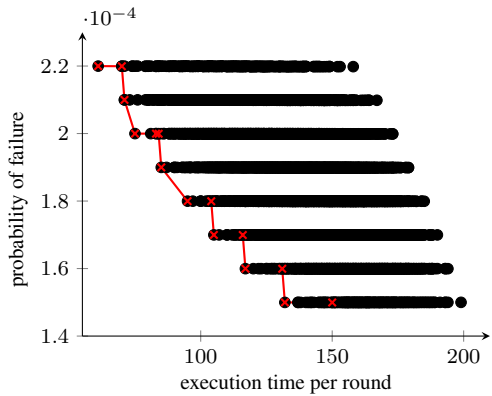
\begin{figure}
	\begin{tikzpicture}
	\footnotesize
        \begin{axis}[
          width=\linewidth, height=20em,
          ymin=0.00014, ymax=0.00023, ystep=0.00005,
          xmin=55, xmax=210, xstep=20,
          xlabel={execution time per round}, ylabel={probability of failure},
          x label style={at={(axis description cs:0.5,0.03)}},
          y label style={at={(axis description cs:0.07,0.45)}},
          axis x line=bottom, axis y line=left]
			
          \addplot[opacity=0.03,only marks] table [col sep=comma]
          {data/air_costprob3.txt};
          
          \addplot[red,opacity=0.8,thick,mark=x,mark size=2.0pt] table [col sep=comma]
          {data/air_pareto3.txt};
        \end{axis}
      \end{tikzpicture}
	\caption{\label{fig:pareto} Configurations Pareto optimal with respect 
		to execution time and probability of failure within two rounds}
\end{figure}

\section{Family-based Analysis of Redundancy Systems}\label{sec:breakinglimits}
The DTMC families generated using the approach we sketched
in the last section enable a family-based reliability analysis.
We illustrate the benefits of such analyses for the PID example and
a large-scale family of a protected aircraft velocity control loop (VCL).
Both model a control loop that operates in \emph{rounds}, i.e.,
starting with initial values of data at the beginning, 
each round is considered to start when the values of data modified in the
last execution of the control loop is fed again as input.
For an analysis using the symbolic probabilistic model checker $\prism$~\cite{prism40}
we then considered the following reliability properties:
\begin{description}
	\item[(pfail)]\!What is the failure probability in $n$ rounds?
	\item[(qround)] What is the maximal number of rounds in which the system can
		guarantee a failure probability below some threshold $\theta$?
\end{description}
The first property is a standard reliability property while
the second is a \emph{quantile}~\cite{BDKDKSW14,BDKL14}.
We rely on an error model that assigns a
fault probability of $10^{-5}$ to each $\simulink$ block.
Throughout presenting the results we abbreviate the redundancy mechanisms
for $\simulink$ blocks as follows: ``--'' stands for no protection, 
``c'' for comparison, ``v'' for voting, and ``s'' for sparing.
All experiments were carried out\footnote{Hardware setup: 
Intel Xeon E5-2680@2.70GHz, 128 GB RAM; Turbo Boost and HT disabled; 
Debian GNU/Linux 9.1} using $\prism$ supporting 
variable reordering techniques~\cite{KBCDDKMM16}.\\

\subsection{Analysis of the PID Controller}
For the analysis of the PID controller, we annotate 
the PID $\simulink$ model of Figure~\ref{fig:pid}
with all the protection mechanisms explained in Section~\ref{ssec:protections}. 
Applying the three redundancy mechanisms thus yields a $\simulink$ model with redundancy 
that depends on the choice of three switch variables selecting the protection combinations.
Hence, after the automated translation of the model using $\simpars$ and $\errorpro$, 
we obtain a DTMC family comprising $4^3=64$ protection combinations.
We then analyzed the \textbf{(pfail)} property with a parameter of $n=10$.
The results provided in Table~\ref{tab:pidpfail} show that the comparison and
voting patterns have higher impacts to protect blocks with a slight advantage for
voting. This is even more apparent when considering the \textbf{(qround)} property, 
also depicted in Table~\ref{tab:pidpfail}. Here, for guaranteeing a failure probability 
below $\theta=3{\cdot}10^{-4}$, a PID that is fully protected by comparison or
voting ``survives'' seven times longer than an unprotected or with sparing protected
PID. For obtaining these results, we performed both, a naive one-by-one analysis 
where every member of the family is analyzed in isolation and an all-in-one analysis. 
However, due to the small number of family members, the all-in-one analysis has only
slight advantages to the one-by-one analysis. 

\subsection{The Velocity Control Loop Model}
To illustrate our approach on a large-scale family of $\simulink$ protections,
we issue a velocity control loop (VCL) of an aircraft model. 
Our model is a simplified version of the aircraft model borrowed from the $\simulink$ 
example set~\cite{aircraft} that itself is based on a long-haul passenger aircraft 
flying at cruising altitude and speed, adjusting the fuel flow rate to control the aircraft velocity.
Figure~\ref{fig:velocity} shows the  $\simulink$ model where the eight blocks amendable 
for protection mechanisms are shaded.
Note that the VCL also includes the PID controller from the previous example.
On each of these blocks we applied the redundancy mechanisms comparison, voting, and sparing
as described in Section~\ref{ssec:simulinkredundancy}, resulting in $4^8=65\,536$ combinations 
of protections. After applying the transformations of Section~\ref{sec:approach_PID},
we first analyzed the generated DTMC family against the \textbf{(pfail)} property.
Here, we used both, an all-in-one approach
performing the analysis on a single family model, and (partially)
a one-by-one approach, checking each combination of protections separately.
We can already observe from the statistics in Table~\ref{tab:analysis} that the generated 
DTMC family model requires symbolic techniques to be analyzed by an all-in-one analysis due
to the massive size of the state space. 
To estimate the sizes of models and analysis times 
for a one-by-one, we considered 655 randomly generated instances of the family 
(i.e., around 1\% of the family members) as analyzing all family members in isolation 
clearly would have exceeded time constraints. 
This fact is already underpinned by the size and and analysis times for the randomly
generated instances, ranging up to %
$1.6\cdot 10^{11}$ states requiring $5\,093.922$ seconds of analysis time.
As even single instances have this magnitude of model sizes and symbolic representations, 
one can already estimate that our approach exploiting redundancy through symbolic representations
is viable.
The one-by-one analysis of all 655 randomly generated family members took around 60.4 hours
such that we estimate the whole analysis for all 65\,536 protection combinations would take
more than 250 days of computation time. This demonstrates a speedup of three orders of magnitude
an all-in-one analysis yields compared to an exhaustive one-by-one analysis.

\subsubsection{Synthesis of Optimal Tradeoff Protections}
To investigate the tradeoff between execution time and reliability,
we measured the impact of protection mechanisms on the execution time
of one round~\cite{more}. Without any protection, each round required 61 time units
to be executed, increased by timings for each protection shown in Table~\ref{tab:time}.
\begin{table}
	\tbl{\label{tab:time}Impact of protections on one-round execution time}
	{\tabcolsep11pt
	\begin{tabular}{@{}lccc@{}}\toprule 
		block&comparison&voting&sparing\\\colrule
		P term &10&15&9\\
		I term &15&23&14\\
		D term &14&22&13\\
		Signal Builder &10&15&9\\
		Fuel Mass&10&15&9\\
		Subtract&12&18&13\\
		Vehicle Mass&10&15&9\\
		vCruise&10&15&9\\\botrule
	\end{tabular}}
\end{table}
As the all-in-one analysis provided the results for each protection combination
in the family model, we hence can easily compute the execution time per round
for each combination and relate them to their reliability properties.
Figure~\ref{fig:pareto} depicts for each family member the probability of failure 
within two rounds (property \textbf{(pfail)}) and its costs in terms of execution time.
The line at the left indicates the Pareto front, which directly yields the
optimal protection combinations when either fixing constraints on the probability
of failure or execution time. Pareto-optimal configurations are shown in 
Table~\ref{tab:pareto}, where the protection combinations correspond to
the chosen protections for the blocks in the order of Table~\ref{tab:time}, i.e.,
the combination ``\makebox[.7em]{-}\makebox[.7em]{-}\makebox[.7em]{c}\makebox[.7em]{-}\makebox[.7em]{s}\makebox[.7em]{-}\makebox[.7em]{-}\makebox[.7em]{-}''
stands for protecting the D term block with comparison and the Fuel Mass block
with sparing.
\begin{table}
	\tbl{\label{tab:pareto}Pareto-optimal protection combinations}
	{\tabcolsep10pt
	\begin{tabular}{@{}lcc@{}}\toprule 
		combination&exec. time&prob. of failure\\\colrule
	\makebox[.7em]{c}\makebox[.7em]{c}\makebox[.7em]{c}\makebox[.7em]{s}\makebox[.7em]{c}\makebox[.7em]{c}\makebox[.7em]{c}\makebox[.7em]{s}&150&$1.4995{\cdot}10^{-4}$\\
\makebox[.7em]{c}\makebox[.7em]{c}\makebox[.7em]{c}\makebox[.7em]{-}\makebox[.7em]{c}\makebox[.7em]{c}\makebox[.7em]{c}\makebox[.7em]{-}&132&$1.4997{\cdot}10^{-4}$\\
\makebox[.7em]{c}\makebox[.7em]{c}\makebox[.7em]{c}\makebox[.7em]{-}\makebox[.7em]{s}\makebox[.7em]{c}\makebox[.7em]{c}\makebox[.7em]{-}&131&$1.5994{\cdot}10^{-4}$\\
\makebox[.7em]{c}\makebox[.7em]{-}\makebox[.7em]{c}\makebox[.7em]{-}\makebox[.7em]{c}\makebox[.7em]{c}\makebox[.7em]{c}\makebox[.7em]{-}&117&$1.5997{\cdot}10^{-4}$\\
\makebox[.7em]{c}\makebox[.7em]{-}\makebox[.7em]{c}\makebox[.7em]{-}\makebox[.7em]{s}\makebox[.7em]{c}\makebox[.7em]{c}\makebox[.7em]{-}&116&$1.6994{\cdot}10^{-4}$\\
\makebox[.7em]{c}\makebox[.7em]{-}\makebox[.7em]{c}\makebox[.7em]{-}\makebox[.7em]{c}\makebox[.7em]{-}\makebox[.7em]{c}\makebox[.7em]{-}&105&$1.6997{\cdot}10^{-4}$\\
\makebox[.7em]{c}\makebox[.7em]{-}\makebox[.7em]{c}\makebox[.7em]{-}\makebox[.7em]{s}\makebox[.7em]{-}\makebox[.7em]{c}\makebox[.7em]{-}&104&$1.7994{\cdot}10^{-4}$\\
\makebox[.7em]{-}\makebox[.7em]{-}\makebox[.7em]{c}\makebox[.7em]{-}\makebox[.7em]{c}\makebox[.7em]{-}\makebox[.7em]{c}\makebox[.7em]{-}&95&$1.7997{\cdot}10^{-4}$\\
\makebox[.7em]{-}\makebox[.7em]{-}\makebox[.7em]{c}\makebox[.7em]{-}\makebox[.7em]{c}\makebox[.7em]{-}\makebox[.7em]{-}\makebox[.7em]{-}&85&$1.8997{\cdot}10^{-4}$\\
\makebox[.7em]{-}\makebox[.7em]{-}\makebox[.7em]{c}\makebox[.7em]{-}\makebox[.7em]{s}\makebox[.7em]{-}\makebox[.7em]{-}\makebox[.7em]{-}&84&$1.9994{\cdot}10^{-4}$\\
\makebox[.7em]{-}\makebox[.7em]{-}\makebox[.7em]{-}\makebox[.7em]{-}\makebox[.7em]{c}\makebox[.7em]{c}\makebox[.7em]{-}\makebox[.7em]{-}&83&$1.9997{\cdot}10^{-4}$\\
\makebox[.7em]{-}\makebox[.7em]{-}\makebox[.7em]{c}\makebox[.7em]{-}\makebox[.7em]{-}\makebox[.7em]{-}\makebox[.7em]{-}\makebox[.7em]{-}&75&$1.9998{\cdot}10^{-4}$\\
\makebox[.7em]{-}\makebox[.7em]{-}\makebox[.7em]{-}\makebox[.7em]{-}\makebox[.7em]{c}\makebox[.7em]{-}\makebox[.7em]{-}\makebox[.7em]{-}&71&$2.0997{\cdot}10^{-4}$\\
\makebox[.7em]{-}\makebox[.7em]{-}\makebox[.7em]{-}\makebox[.7em]{-}\makebox[.7em]{s}\makebox[.7em]{-}\makebox[.7em]{-}\makebox[.7em]{-}&70&$2.1994{\cdot}10^{-4}$\\
\makebox[.7em]{-}\makebox[.7em]{-}\makebox[.7em]{-}\makebox[.7em]{-}\makebox[.7em]{-}\makebox[.7em]{-}\makebox[.7em]{-}\makebox[.7em]{-}&61&$2.1998{\cdot}10^{-4}$\\\botrule
	\end{tabular}}
\end{table}
Similar to the plain PID example, we observe that protection with comparison
has great impact on the probability of failure, appearing in most of
the Pareto-optimal combinations. Voting does not show up in optimal
combinations (although possibly as good as comparison by means of 
gained fault tolerance) due to its comparably high 
execution time (see Table~\ref{tab:time}). 
Sparing does not reduce the probability of failure  as much as comparison and voting,
but has good timing characteristics such that it appears at some
occasions in the Pareto-optimal combinations.

\paragraph{Remarks on Limitations.}
Whereas impossible for the quantile property, the standard reliability
property asking for the probability of failure within a fixed
number of rounds could also be evaluated using simulation-based
approaches. However, due to  the relatively 
small probability of faults in each P, I, or D term, 
we were not able to perform statistical model checking with
sufficient confidence in either the PID and VCL model.
We also performed analyses for the quantile property
\textbf{(qround)}, but due to the size of the models already single instances
of the DTMC family could not be analyzed within a week of computation such 
that we dropped an exhaustive study of this property for the VCL model.\\[.1em]

\subsubsection{Further Techniques Applied}\label{ssec:opti}
As usual, automatically generated models for $\prism$ do not admit
a good variable ordering for their concise symbolic representation via MTBDDs
and thus require post-processing steps to be amendable for a formal analysis
(see, e.g., \cite{KBCDDKMM16,CDKB16}). Within our tool chain, things
were actually worse as even single instances of the $\errorpro$ generated 
DTMCs could not be built without either running out
of memory or taking several days before interrupting the building
process. To enable an analysis of our models, we had to apply the
following post-processing steps on the generated VCL model.

\paragraph{Reset Value Optimization.} 
Thanks to the control- and data-flow models in the DEPM meta model, we used standard
graph algorithms applied for each data element to determine those control-flow 
locations where the data is never read before it is written again. 
For the minimal control-flow locations (with respect to the control-flow order) we 
reinitialized the value of the data storage to the value it has at the initial control-flow location.
The ratio behind this optimization is that the state-space is reduced by joining
naively bisimilar states where data values do not have impact on the
future behaviors.

\paragraph{Iterative Variable Reordering.}
We exploited further the sensitivity of MTBDD represented models 
to its variable ordering. For this, we successively built subfamily
models, iteratively adding protection mechanisms to each
block and performing variable reordering \cite{KBCDDKMM16}
to determine a suitable variable order.  %

\section{Discussion and Further Work}\label{sec:conclusion}
We proposed to use family-based approaches for the modeling and analysis
of redundancy models to overcome limitations imposed by the
combinatorial blowup that arises when protecting system components.
To illustrate the approach, we presented a tool chain that enables reliability analysis
of $\simulink$ models with redundancy. 
In future, we aim at further automating the handcrafted steps in this tool chain,
e.g., incorporating the optimizations done in Section~\ref{ssec:opti} into $\errorpro$.

Note that although presented in the specific setting for $\simulink$ designs,
our approach is applicable to many redundancy system models. 
In particular, the tool chain we present enables to use many base-line 
models supported by the tool used for 
the automated translation towards DTMC families, e.g., by $\errorpro$~\cite{morozov2015errorpro},
and the full analysis power of properties that can be checked for DTMCs
using, e.g., by symbolic probabilistic model checkers such as $\prism$~\cite{prism40}. 
\paragraph{Acknowledgments.}
This work is supported by the DFG through
the Collaborative Research Centers CRC 912 (HAEC) and
TRR 248 (see {\footnotesize\url{https://perspicuous-computing.science}}, project ID 389792660),
the Cluster of Excellence EXC 2050/1 (CeTI, project ID 390696704, as part of Germany's Excellence Strategy),
the Research Training Groups QuantLA (GRK 1763) and RoSI (GRK 1907), 
projects JA-1559/5-1, BA-1679/11-1, BA-1679/12-1, and the 5G Lab Germany.

\bibliographystyle{abbrvnat}
\bibliography{main-short}

\end{document}